\documentclass[preprint, review]{elsarticle}
\usepackage{epsfig,endnotes}
\usepackage{algorithm}
\usepackage{algorithmic}
\usepackage{graphicx}
\usepackage{amsfonts,amsmath,amstext}

\journal{Physics A}
\begin{document}
\begin{frontmatter}


\title{\Large \bf Explore what-if scenarios with SONG: \underline{So}cial \underline{N}etwork Write \underline{G}enerator}

\author[tef]{ Vijay Erramilli\corref{cor1}}
\ead{vijay@tid.es}

\author[tef]{Xiaoyuan Yang}
\ead{xiaoy@tid.es}
 
 \author[tef]{Pablo Rodriguez}
\ead{pablorr@tid.es}

\cortext[cor1]{Principal corresponding author}
\address[tef]{Telefonica Research, via Augusta, 177E, Barcelona 08021, Spain}



\begin{abstract}
Online Social Networks (OSNs) have witnessed a tremendous growth the last few years, becoming
a platform for online users to communicate, exchange content and even find employment. 
The emergence of OSNs has attracted researchers and analysts and much data-driven research
has been conducted. However, collecting data-sets is non-trivial and sometimes it is difficult
for data-sets to be shared between researchers. The main contribution of this paper is a
framework called SONG (Social Network Write Generator) to generate synthetic traces of write 
activity on OSNs. We build our framework based on a characterization study of a large 
Twitter data-set and identifying the important factors that need to
be accounted for. We show how one can generate traces with SONG and validate it by comparing against real
data. We discuss how one can extend and use SONG to explore different `what-if' scenarios.
We build a Twitter clone using 16 machines and Cassandra.
We then show by example the usefulness of SONG by stress-testing our implementation. 
We hope that SONG is used by researchers and analysts for their own work
that involves write activity.   
\end{abstract}
\end{frontmatter}

\section{Introduction}
\label{sec:intro}
\vspace{-2mm}
OSNs have increased in popularity the last few years and they attract
hundreds of millions of users. OSNs differ from traditional web applications
in atleast two respects: they serve highly personalized content and
have to deal with non-traditional workloads \cite{user-imc,osnnwk-imc}. 

Complimenting this rise of popularity has been the vast amount of research
done on OSNs -- from analyzing social communities\cite{leskovec, imc-moon,kdd-jon}, to studying
presence of social cascades in OSNs \cite{user-imc,user-fbook,gao,cha-2008,cha-2009} 
and solving problems of large scale OSNs \cite{jmpujol}. Most of this work is aided by real
datasets collected by elaborate crawls on OSNs. However gathering datasets is hard for 
multiple reasons - some OSNs like Facebook have high privacy settings making it hard to obtain
a proper sample. OSNs like Twitter have
caps on the number of API calls that can be made\footnote{http://apiwiki.twitter.com/Rate-limiting}, making
a large crawl a long and elaborate affair. And sometimes, data that has been collected 
cannot be released due to legal reasons \cite{moon}. 
Lack of data inhibit researchers and analysts to properly conduct research.
In addition, even when a sample is collected, the sample only captures characteristics
present at the time of collection, limiting the scope of using the sample. 

The main contribution of this paper is SONG - a framework to generate synthetic and realistic 
traces of write activity of users in OSNs. We focus on modeling writes as from a system perspective,
writes have more impact than reads (issues of consistency etc.) and a read is performed whever a write event occurs.
In order to generate synthetic traces, 
we treat the problem of generating writes as a classical time-series modeling problem of a count process,
with the number of writes in given time interval representing counts.
We build a model
of write activity of users that can span long time scales - days as well as short time scales - seconds.
In cases where researchers and analysts 
have no data, they can use SONG to generate traces that are realistic. And in cases where
some datasets are available, researchers and analysts can use SONG to estimate parameters from
the existing datasets, and generate new traces to explore different \emph{what-if} scenarios, 
including forecasting, benchmarking, capacity planning, performance analysis and 
effects of flash crowds.

In order to understand what factors effect write activity and hence 
are necessary to consider in any framework in order to 
generate realistic traces, we characterize different user-centric properties of data
collected by a large crawl of Twitter along with associated writes (in the form of tweets) 
\cite{pujol-2009}. We confirm some 
findings that have been reported before -- the presence of diurnal trends \cite{gao}, a small subset of users
have high activity; the activity of users follows a skewed distribution and 
geographical properties of users; most users are from the US\cite{twitter}.  We also report new ones  -- 
to the best of our knowledge, inter-write activity of users is distributed log-normally in time, a finding that
is closer to observed online behavior of humans \cite{stouffer-2005}. We also observe writes do not show self-similar behavior.

SONG is based on two components - the first component is responsible for modeling and
generating time-varying trends that operate at longer time scales (diurnal variations). 
The second component is responsible for modeling activity at short 
time-scales (seconds - a couple of hours). The model we pick is simple, intuitive 
and fits our purpose.  We provide guidelines to generate traces
when no data-sets are available using standard off-the-shelf methods. 
When data is available, we show how one can estimate
different parameters that are used by SONG. We validate SONG using real data and show 
that traces generated by SONG match real data closely. 

As statistician G.E. Box put it: ``All models are wrong, some models are useful". To show the efficacy
of SONG on a real system, we implement a clone of Twitter using Cassandra \cite{cassandra} on 16 commodity machines.  
 We validate the trace generated by SONG against the real data set, by
 monitoring system-specific metrics like CPU activity, IO activity and network traffic produced in the back-end. 
 We then show by example the usefulness of SONG - by considering a simple 
 yet crucial scenario - benchmarking the system and uncovering bottlenecks.  
\vspace{-2mm}

\section{Related Work}
\label{sec:relwork}
\vspace{-3mm}
The main contribution of this paper is a framework for generating 
time-series of activities (writes) of users in OSNs. Towards this end,
we build up on various concepts from previous work done in this area. We 
compare and contrast our work with related work in this section. 

Much work has been done on characterizing and understanding information propagation 
in social networks \cite{cha-2009,cha-2008,oh08,infopath,kempe}. However, most of this 
work deals with the spread of \emph{content} (in an OSN like 
Flickr \cite{cha-2009, cha-2008}) and what factors are responsible for 
information cascades to form. Our work is different as we deal with 
modeling of user activity over time; writes in an OSN and how such 
a model can be useful. In addition, we focus on modeling an aggregate of users,
instead of focusing on individual users or a specific community of users.

The closest to our work is the work on understanding user 
interactions in OSNs \cite{user-fbook, gao,walter-wosn,visw-wosn2009,moon}.
The authors of \cite{user-fbook} use a crawl of Facebook to extract a 
social-interaction graph and show that such a graph deviates from the 
social graph and how these differences can manifest themselves in 
gauging the performance of social applications running on these graphs. A 
similar message was delivered by the authors of \cite{walter-wosn} where they 
advocate using a more `dynamic' view of the social graph, rather than a 
static friendship graph for analysis. The authors of 
\cite{gao} investigate the posting behavior of users in OSNs and show that users 
exhibit strong diurnal patterns, something we observe as well in our datasets. In 
addition the authors show that the posting behavior follows a 
stretched-exponential distribution. In our datasets, we see that user 
posting behavior (tweets) follows a log-normal distribution and is more 
in line with human response dynamics \cite{stouffer-2005}. We use these properties
while building our model. Finally the authors of \cite{visw-wosn2009} study 
the evolution of social links over time, by studying user interaction. 

Our work can be seen as complementing recent work on characterizing user 
workloads on OSNs by analyzing clickstream data \cite{user-imc, osnnwk-imc}.
These works primarily focus on studying the impact of OSNs on the network, by 
studying how users interact with different OSNs. Our work is markedly different
as we focus on modeling interactions in an OSN and we believe that our model
can be incorporated in a more comprehensive model of user behavior. 

Finally, we use data gleaned from the Twitter network for our work,
and we find similar behavior as has been observed \cite{twitter,www-moon}. To the 
best of our knowledge, this is the first work dealing with modeling of 
interactions (writes) in an OSN with the hope that researchers and analysts alike can
use the model to generate synthetic traces for their own work.
\vspace{-2mm}

\section{Data}
\label{sec:dat}
\vspace{-3mm}

Our dataset consists of a crawl of Twitter \cite{twitter} conducted between
Nov 25 - Dec 4, 2008. The social graph that we crawled
has $2,408,534$ nodes and $38,381,566$ edges (although
Twitter has directed edges, we report total edges and use edges as undirected 
unless otherwise stated). The crawl was done by a standard BFS. 

\begin{figure}[t]
\centering
\includegraphics[width=2.0in]{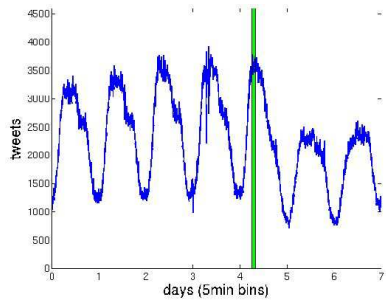}
\vspace{-3mm}
\caption{Tweets for 1 week, 5 minute bins}
\label{fig:tweets_1week}
\end{figure}

We also collected traffic on Twitter in the form of `tweets' (total 12M tweets) 
generated by the 2.4M users by using the Twitter API\footnote{http://apiwiki.twitter.com}. 
From the tweets, we mined the following relevant 
fields: {\em tweet\_id}, {\em timestamp}, {\em user\_id}, {\em location} and 
{\em content}. This information allows us to 
get traffic information for every user and
how this traffic is distributed across users.
Approximately 25\% of the population (587K) generated at least one tweet, 
for the rest of the users the Twitter API did not return tweets in the 19-days period 
under examination.  Given the broadcast nature of Twitter (a tweet is sent to all the 
followers of a user), the 12M tweets lead to an excess of 1.7B messages actually 
traversing the system. An initial study of our data (and borrowing the terminology used in \cite{twitter}) 
revealed the presence of \emph{broadcasters} - users or automatic programs (like `cnnbrk') that publish 
tweets at high frequency and normally have many followers as well as \emph{miscreants} or spammers
who also publish tweets at a high rate but have few or no followers. 
As we are interested in modeling legitimate users, we filter out these spammers by using a simple
method - all users who have more than 80\% of the maximum number of tweets (during the 19 days)
and have less than the median number of followers are filtered.  In our dataset, 
the maximum number of tweets by a single user was 1000, with mean and median number of tweets per
user being 21.2 and 8 respectively, with the standard deviation being 35.08. 
The median number of followers is 8, the mean is approx. 64. We therefore classified all
users who have more than 800 tweets and less than 8 followers as spammers and removed them.
A manual inspection revealed all high-rate 
spammers were removed. It is possible low-rate spammers are present in the dataset, but they do not 
skew our modeling process. In addition, we also removed users who tweet only once during the 19 day 
period. We ended up with 346424 users. 
There exist larger data-sets of Twitter \cite{www-moon}, however the traffic information collected 
is different (traffic pertaining to trending topics) and does not consist of the 
tweets of individual users. To the best of our knowledge, this is the largest dataset 
with user activity (tweets). 
In this paper, we present results from 1 week (shown in Fig.~\ref{fig:tweets_1week}).
For further evaluation, we also consider a trace of 2 hours labeled as the `busy-hour' trace where 
the mean and variance are fairly stable. (highlighted in green). 

\textbf{Limitations with the data:}
While we are aware that unbiased sampling methods like Metropolized sampling method (MRW) exist \cite{unbiased}, 
the size of our dataset when collected suggest that a large portion of the social graph 
was captured. When we collected the tweets of individual users, we noticed tweet ids were non-contiguous. 
This implies the API missed returning some tweets. A cursory look at the id distribution in our dataset
reveals the API could have lost upto 60\% of the tweets (although uniformly distributed), 
however, we do not have additional information to confirm this number. 

\vspace{-2mm}

\section{Characterization}
\label{sec:char}
\vspace{-3mm}

In order to develop a framework to generate synthetic traces that are realistic,
we need to first understand what facets of data are important enough to consider. 
In addition, if we want the framework to be beyond merely curve-fitting; descriptive, 
we need to consider processes that are generative. For this, we study properties
of the dataset we collected to guide our design.  

\begin{figure*}[tbp]
\centering
\begin{tabular}[2]{cc}
\includegraphics[width=2.0in]{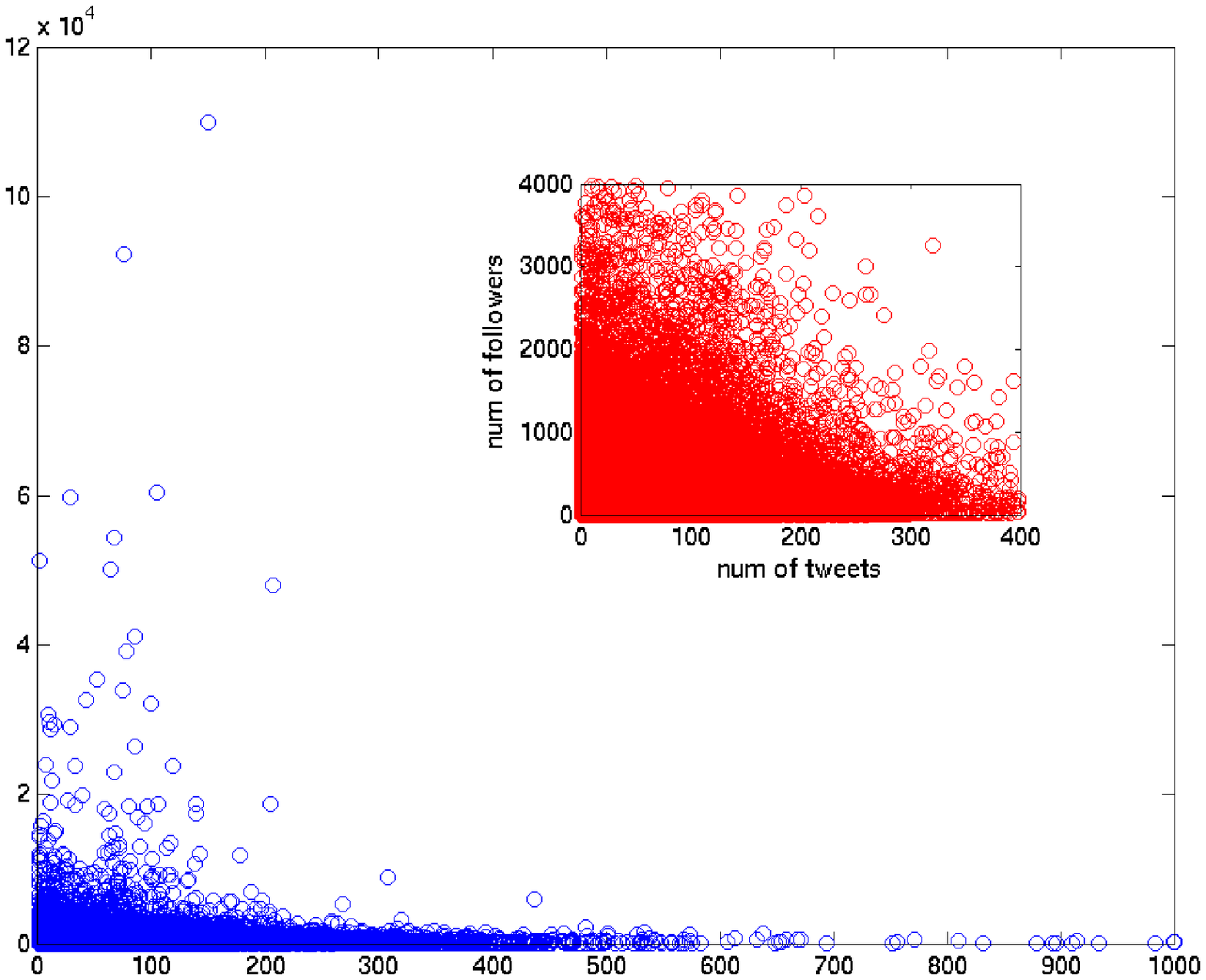}
&
\includegraphics[width=2.0in]{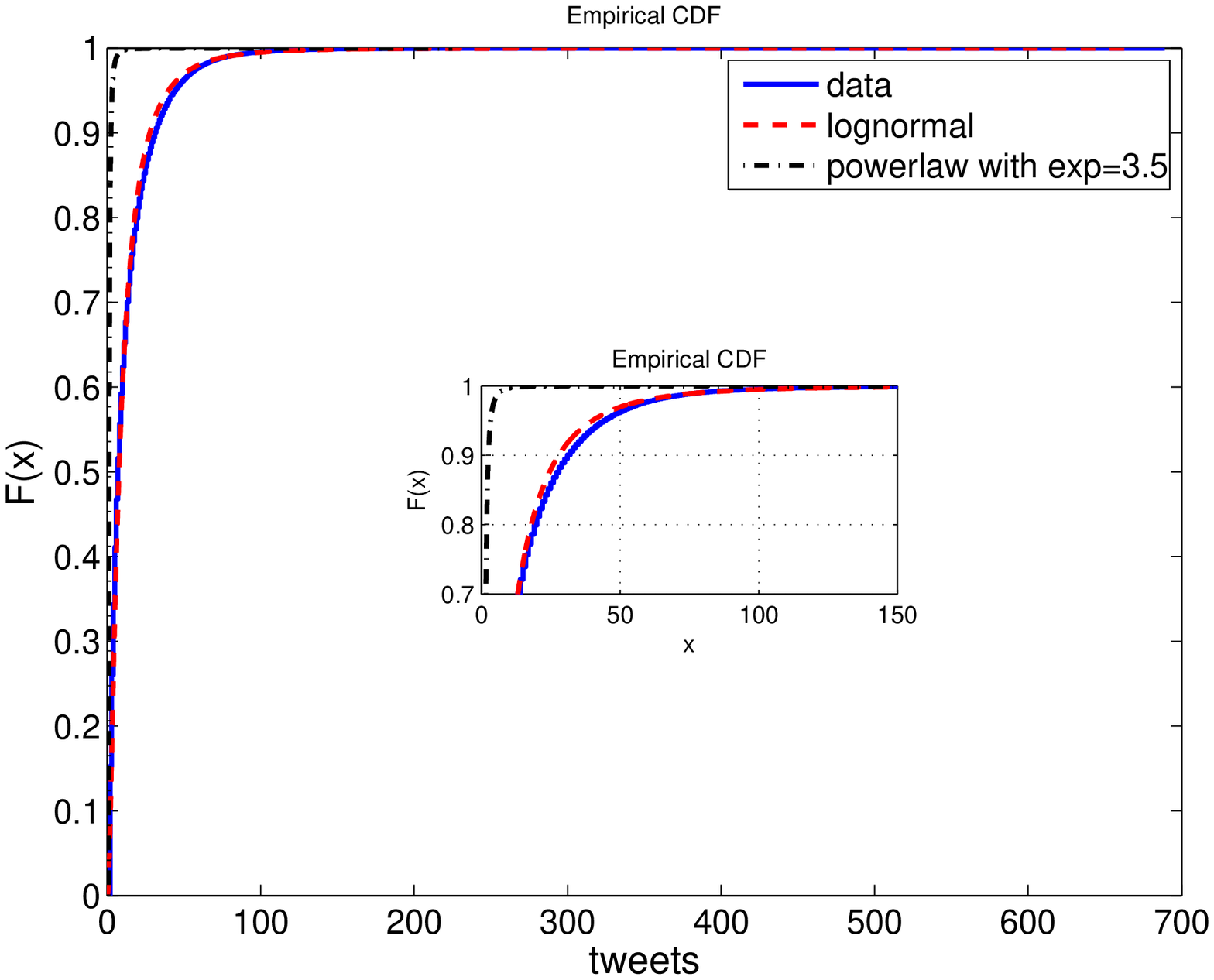}
\\
(a) & (b)
\end{tabular}
\vspace{-3mm}
\caption{ (a) Number of followers vs writes, (b) CDF of writes per users}
\label{fig:tweetprop}
\end{figure*}

\subsection{User Properties}
\label{subsec:user}
\textbf{Diurnal Properties}: From Fig.~\ref{fig:tweets_1week}, we see that users follow a strong diurnal pattern
with respect to writing (tweeting) and the number of writes are bursty, and is similar to traffic patterns observed in IP networks. 
We need to account for this trend in SONG.

\textbf{Correlation between degree and rates}: Do users write more because they are high-degree nodes or
are they high degree nodes because they write more? To answer this question, we first plot the number of 
followers against the number of writes for each user in Fig.~\ref{fig:tweetprop} (a) and we note there is 
little or no discernible correlation (a similar observation has been reported before \cite{www-moon}). We observed 
the same when we plotted number of writes against the number of friends. This could be the artifact of the system 
we consider - Twitter, that lets
users follow or be followed without an explicit social relationship. For our purposes we do not consider
incorporating the degree of a user (number of followers/friends) in our model explicitly, although we show later how 
SONG can be extended to incorporate this. 

\textbf{Distribution of activity between users}: Not surprisingly, users show high variance when it comes 
to the number of writes they generate (also observed in \cite{www-moon}) (Fig.~\ref{fig:tweetprop} (b)). We try 
fitting the distribution with a power-law ($ p(x) \propto x^{-\alpha}$, using the method described 
in \cite{clauset}) and a log-normal distribution 
($p(x) = \frac{1}{x\sigma\sqrt{2\pi}}e^{-\frac{(\ln x - \mu)^2}{2\sigma^2}}$) and we note an excellent fit with the log-normal 
distribution (parameters 2.05, 0.9921). We note that a power-law is more appealing as it is more parsimonious, but we do not see a good fit. 
How does this distribution arise? We know that log-normal distributions arise
by multiplicative processes \cite{mitzen}, however to fully characterize this underlying multiplicative process is beyond the 
scope of this current paper. We incorporate distribution of activity in SONG.

\textbf{Inter-write distribution for individual users}: An important property to understand is when do 
individual users write; the inter-write distribution of users. For this, we focused on 
the top 10000 users in terms of the number of tweets and found the best fit for the inter-write distribution 
is the log-normal distribution, which is closer to what has been reported
for email communication \cite{stouffer-2005} etc. The Kolmogorov-Smirnov test on the log-normal fits yielded a positive 
result for 98.1\% of the users, while it was 32\% for the fits with the exponential distribution.  
However we make a conscious decision to model aggregate of users, and the aggregate inter-write distribution
follows an exponential distribution \cite{albin}.

\textbf{Geographical Properties}: To construct a framework from \emph{first-principles} where different parameters
of a model have a physical interpretation, using geography as input to a model is highly appealing \cite{lj-rbf,twitter}.  
After basic preprocessing of our data we obtained 187K different
locations for the 996K users where the location
text was meaningful. 
After filtering, we obtained standardized locations for 691K users. Of these users, 60.2\% belong to the US 
with UK coming a distant second 
with 6.2\% of the nodes. Of the users we consider (~346K), around 82\% of them belong to the US. As
we do not have sufficient information, we do not consider geography or locality properties in our framework
and leave it for future work.

\begin{figure}[t]
\centering
\includegraphics[width=2.2in]{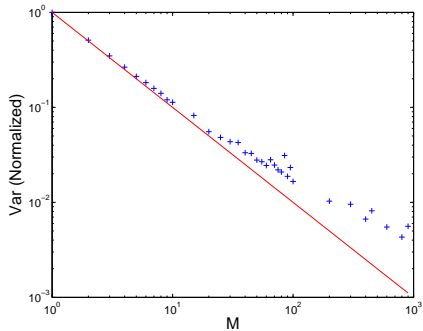}
\vspace{-3mm}
\caption{Variance time plot for busy hour (2hr)}
\label{fig:tweets-ss}
\end{figure}

\subsection{Statistical Properties}
\label{subsec:stats}

\textbf{Test for independence in arrivals}: OSNs are known to experience information cascades, where content 
becomes popular by word of mouth and gets propagated on social links \cite{cha-2008,cha-2009}. This would imply 
that independence of arrivals (at smaller time scales) does not hold. 
A generic test for independence in write arrivals is autocorrelation  -- low values at different 
lags signify little or no correlation. The autocorrelation value steeply decreases for our set 
of arrivals, with lag=1 having a low value of 0.11 - implying statistical independence. This can be explained
as we work at an aggregated level (aggregated over all users), and correlations can be washed out. 

\textbf{Test for self-similarity in write counts}: Self-similar behavior in the time series 
of writes implies that there is no `natural' length for 
a burst of writes; the bursts appear on a wide range of time scales.  From an engineering 
perspective, self-similar behavior in traffic can have 
a profound impact on the design, control and analysis of such systems.  
A lot of work has been in observing self-similar behavior in IP networks \cite{leland,PaxsonF95}, modeling \cite{PaxsonF95} as well as performance
analysis with self-similar traffic \cite{ashok}. It is important to account for self-similar behavior in a model, if it is present. 
A simple visual test for self-similar behavior is the variance-time plot. If we define $X = (X_t:t=0,1,2..)$ to be a stochastic process that represents the
time-series we deal with, then we can create a new time-series  $X^{(m)}$ for each $m=1,2,3..$ by averaging the original series $X$ over non-overlapping blocks 
of size $m$. For self-similar processes, the variances of aggregated process $X^{(m)} (m = 1,2,3..)$ decrease linearly in log-log plots with respect to 
increasing $m$ with slopes arbitrarily flatter than -1. However, for normal processes, this slope is close to -1 \cite{PaxsonF95}. 
Strictly speaking, a rigorous analysis of the presence/absence of self-similar behavior needs datasets that span more time-scales, we don't have such data-sets.
However, we still consider the `busy-hour' trace (as it displays stationarity) and plot the variance-time plot (with aggregated variance  
over 3 orders of magnitude) in Fig.~\ref{fig:tweets-ss}.  We note that the slope is close to -1, showing a lack of self-similar behavior. 

In this section, we have tried to enumerate properties of data that a framework has to consider. In the next section, we describe our 
framework SONG that build up on what we learned in this section. 

\vspace{-2mm}

\section{Framework and Trace-Driven Evaluation}
\label{sec:model}
\vspace{-2mm}

\begin{figure*}[tbp]
\centering
\begin{tabular}[2]{cc}
\includegraphics[width=2.3in]{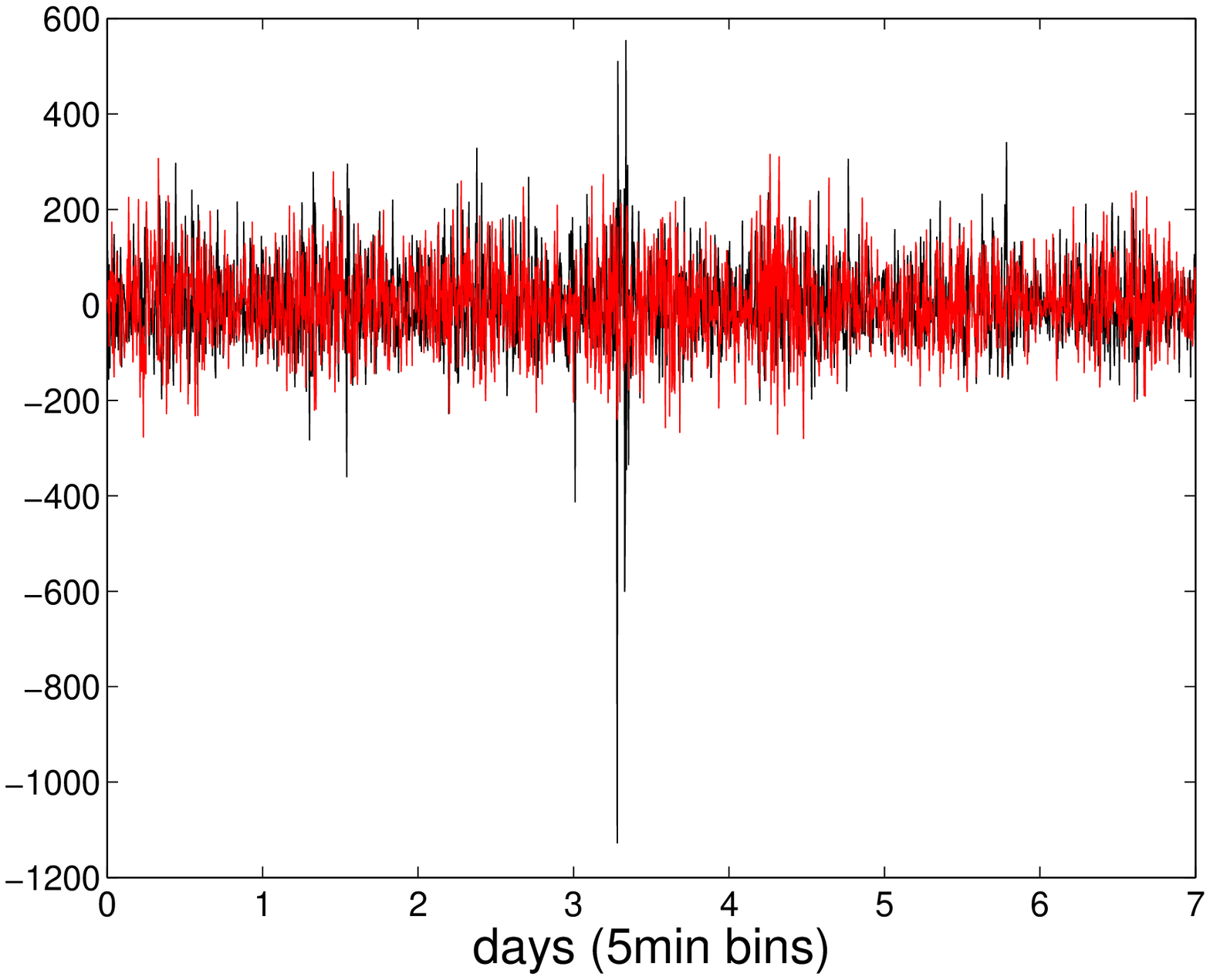}
&
\includegraphics[width=2.4in]{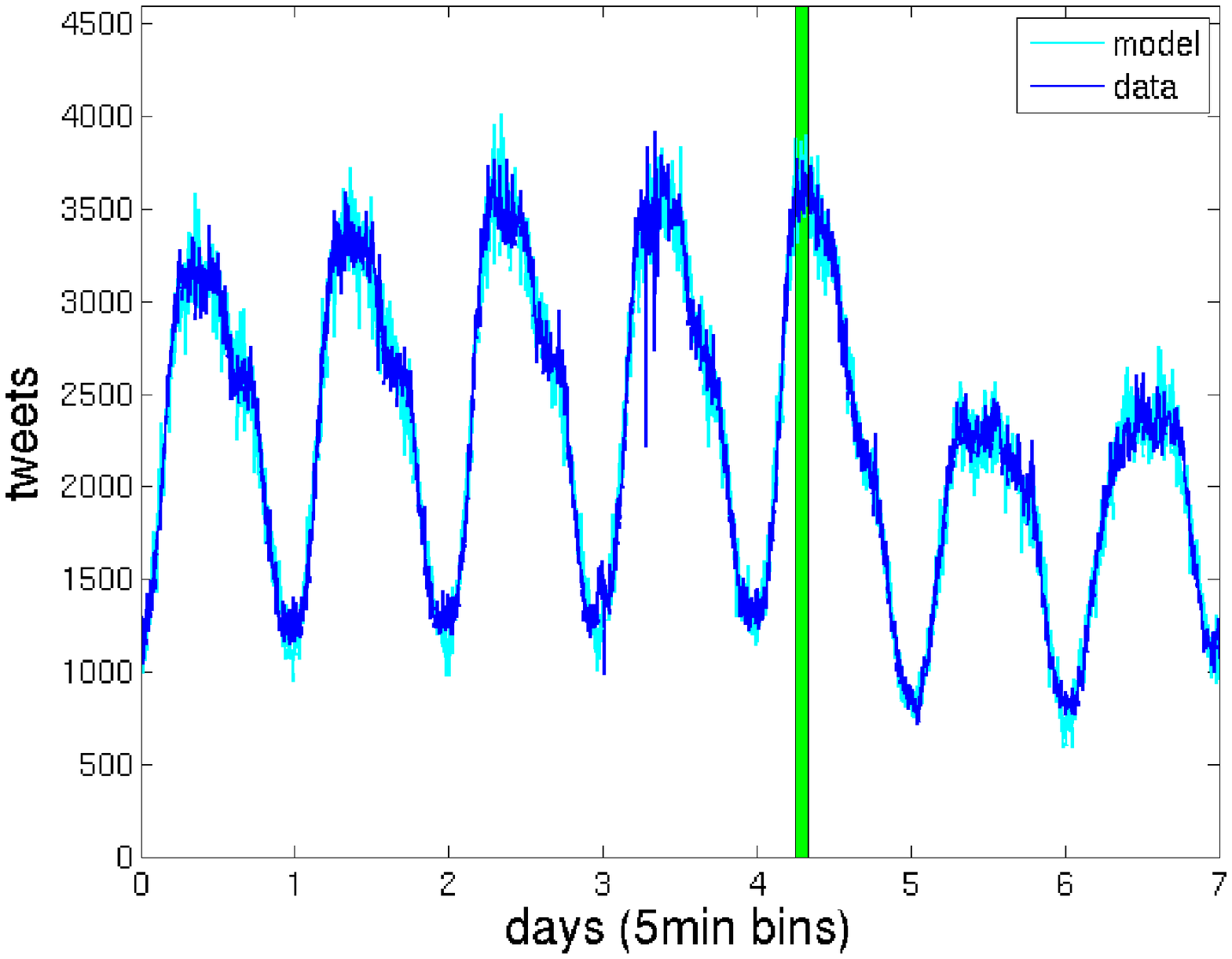}
\\
(a) & (b)
\end{tabular}
\caption{ (a) Noise model, (b) Model vs Data}
\label{fig:tweetsfft}
\end{figure*}

In this section, we first describe the model that forms the basis of SONG to generate 
synthetic traces. We then evaluate the framework by using our datasets.

\subsection{SONG - Methodology}

Let $X_i(t)$ denote the number of writes produced by user $i$, $1\leq i \leq N$ with $N$ the total number of users 
at a time instant $t$, where $X(t) = \sum_{ \forall i}  X_i(t)$.  The time can vary from seconds to weeks, we
focus on one week for modeling purposes, with the basic unit being 1 sec. The description $X(t), \forall t$ 
gives the time series aggregated over all users. We start with the observation that we need to account for two
different time-scales - the first time scale spans multiple hours or days and we note the presence of diurnal 
trends (from Sec. \ref{subsec:user}). The second time scale spans seconds to a couple of hours where the mean 
and the variance are fairly stable. For the first time scale, we can have a model for the mean $m_t$ 
of the time series that varies with time in a predictable way. For the second time scale, we can have 
a stochastic component. The model then is
\begin{equation}
X(t) = m_t + \sqrt{am_t}W_t
\label{eqn1}
\end{equation}

where $m_t$ is function of time and 
$W_t$ is a stochastic component which can be a zero-mean, finite variance process and $a$ is
a parameter called `peakedness' (with the same units as $X(t)$) 
that accounts for magnitude of fluctuations.
This model has been used before for modeling backbone traffic in IP networks with success \cite{roughan},
and is appealing to use in our context for the following reasons: it\footnote{We abuse the notation here as the original model was
intended to model a continuous count process, we are using it to model a discrete process.} accounts for diurnal variations as well
as short time scales, allows for using an infinite variance process like Fractional Gaussian Noise (FGN)
to account for self-similar behavior \cite{fgn} or a finite variance process, what we note in our data. 
The model is also intuitively appealing as it captures the effect of multiplexing many sources, in this 
case, users. We now describe how to model different parameters and generate traces.

\textbf{Modeling $m_t$:} In order to model the time varying mean, we can use a set of basis functions 
(Fourier, wavelet or principal components) that capture the main frequency components of the data. 
Ideally, one would associate the presence of different frequencies in the diurnal trend to factors like
geography (users in different time zones becoming active during different times), however a detailed
characterization of this is not possible as we lack the data. We do however have the following facts 
to guide us. First of all, the mean of the time series can be fixed. This is related to the number of
users in the OSN. Then we can generate time-varying waves (like sinusoids) with a given amplitude and 
frequency - for instance the largest amplitude wave normally has a periodicity of 24Hrs. 
The next largest waveform has a periodicity of
12Hrs etc. A linear combination of these waves can form realistic diurnal patterns.
If data is available, then one can directly use a time decomposition method of choice (Fourier, wavelets etc)
and pick the adequate the number of components.

\begin{table*}[t]
\centering
\begin{tabular}{| p{4cm} | p{4cm} |  p{4cm} | }
\hline
 Methodology & Empirical Data & Off-the-shelf \\ \hline
1. Generate $m_t$ & Use Fourier/wavelet transforms & Generate and select basis functions \\
  &  and select appropriate number of coefficients &    to represent diurnal variations \\ \hline 
  2. Generate $W_t$ & Use WGN/FGN by estimating parameters from data & Use WGN/FGN with self-selected parameters \\ \hline
  3. Pick $a$ & Estimate from data & Pick a suitable value \\ \hline
  4. Generate time series of interactions & Use suitable distribution from data & Use a suitable distribution \\ 
    and assign users to interactions  \\ \hline
\end{tabular}
\caption{Methodology to generate synthetic traces}
\label{meth}
\end{table*}

\textbf{Modeling $W_t$, $a$ :} 
$W_t$ can either be a zero mean, finite variance process (like white gaussian noise) or 
a process with infinite variance (like FGN). Once we decide on $W_t$, we need to fix $a$. 
As described earlier $a$ represents intuitively refers to how the unit variance (from $W_t$) should 
scale at a given time $t$. For higher values of variance, a higher value of $a$ should be given and 
vice versa.
If we have 
data, we can estimate $\hat a$ as $Var(z_t)$, where $z_t = \frac{X(t)-\hat m_t}{\sqrt{ \hat m_t}}$, given we
have $\hat m_t$ from the step above and $x_t$ is the data.

\textbf{Generating Traces:} We can use Eqn.~\ref{eqn1} to generate the time series of interactions, but we still have 
to assign these to actual users. We can use a distribution that characterizes the number of interactions
per user 
and can perform inverse sampling \footnote{http://en.wikipedia.org/wiki/Inverse\_transform\_sampling} to assign
interactions to users. This distribution can factor in degree distribution (if there is correlation between
degree of a user in a social graph and the number of interactions the user produces), or use distribution
weighted by the pagerank of users (as shown recently in \cite{www-moon} to be important). 
The entire methodology is presented in Table~\ref{meth}.

\subsection{Validating SONG with real data}
In order to validate SONG with real data, we proceed as follows. We pick the 
data from \emph{previous} week. 
We use the Fourier series to represent the time series. 
After performing the fourier decomposition, we note that the top 10 coefficients account
for 81\% of the total variance. We therefore use only these 10 coefficients and generate
a the diurnal trends $\hat m_t$. Note that the first component is the mean of the time series.

In order to choose an appropriate noise model for $W_t$, we go back to our data
and extract the residual noise by subtracting $\hat m_t$ from the data of the week in consideration. 
The residual noise is shown (in black) in Fig.~\ref{fig:tweetsfft} (a). Visual inspection 
as well as further study using the QQplot and
the Anderson-Darling(AD) test reveals the noise to be close to a normal distribution. Hence we use a zero-mean, 
and a unit variance process for $W_t$. 
We estimate $\hat a$ to obtain $\hat a=3.4$. In the small dataset we have (19 days), the 
value of $\hat a$ is stable, although we need more data to establish any property. When we regenerate the noise
process (second term of Eqn.~\ref{eqn1}) using a zero mean, unit variance $W_t$, the estimated $\hat a, \hat m_t$,
we get a close match (in red) as can be seen in Fig.~\ref{fig:tweetsfft} (a). We then generate the time-series of
writes using Eqn.~\ref{eqn1}.

From Sec.~\ref{subsec:user}, we learnt that the tweet counts per user follow
a log-normal distribution. We therefore use this distribution to assign 
tweets to users. The result is presented in Fig.~\ref{fig:tweetsfft} (b), and we note a close fit 
to real data, reproducing long-term trends (dirunal trends) as well as short term trends. 
 
\textbf{Limitations of SONG:}
One possible route to build a framework is to model behavior of individual users in the social network,
generate interactions of each user, and then aggregate these interactions. This is an appealing option
as one can directly consider user-centric phenomena like information cascades as well as dynamics of 
interactions between users \cite{visw-wosn2009}. 
This however, has the problem of making a model much more complex, with more parameters to be considered. 
We chose to work at a level where users are aggregated and have a much simpler model for our purpose. 
At present, we don't account for cascades explicitly and this is left for future work. 

Although SONG is a fairly general and flexible framework, our evaluation has been done
based on a dataset from Twitter - a pub-sub system and some of the results presented 
may not extend to other popular OSNs like Facebook, Orkut etc. However, we argue that
even for other OSNs, the write activity is broadcasted to 
one's neighbors (e.g. wall postings, `I likes' in Facebook), and would tend to follow similar patterns.  
We focus solely on write patterns in OSNs as obtaining read patterns is near impossible by crawls. 
If read activity is closely correlated to write activity, then we can produce reads as well.

 \subsection{What-if scenarios with SONG}
\label{subsec:whatif}
We present three scenarios where SONG can be useful.

\textbf{Forecasting/Capacity Planning:} Consider the scenario where the user base of an OSN 
doubles overnight, leading to an increase in interaction rate. This can be easily 
modeled by varying $m_t$ in the model, generating traces and studying them. 
Some OSN architectures use message queues \cite{url:twit}, that can be modeled as a network of queues
and the performance of the system can be gauged under different scenarios using trace-driven simulations
using traces generated by SONG.

\textbf{Traffic Analysis:} OSNs can face `anomalous' events like flash crowds and high 
trending topics (like `Michael Jackson' etc.) \cite{www-moon} that change the profile of traffic.
In order to study such
phenomenon, one can generate `normal' traffic using SONG to compare or use SONG to generate traffic with such
events. A simple extension to the model we use in Eqn~\ref{eqn1} is:
$X(t) = m_t + \sqrt{am_t}W_t + I_t$ where $I_t$ is an indicator function and 
can be used to inject high traffic at different time intervals to mimic flash-crowds or high loads. 
This can be used for benchmarking as well.

\textbf{Rise of popularity of OSNs in different geographical areas:} 
Consider the case where an OSN is currently popular in the US and suddenly becomes popular in a European
country. This popularity manifests itself as an increase in the number of interactions from a certain
geographic area of the world, with a different temporal profile. This can be handled by producing
traces for one time-zone, time-shifting this set of traces to produce a new set 
and aggregating the two.

\vspace{-2mm}

\section{Exploring what-if scenarios}
In previous section, we described our framework SONG in full detail, validated the framework with
real data, discussed the limitations and briefly discussed some scenarios where SONG can be used.
In order to show the effectiveness of SONG for testing real systems, we 
implement a Twitter-like system and test this system against different traces generated by SONG.

\begin{figure*}[tbp]
\centering
\begin{tabular}[2]{ccc}
\hspace{-0.7cm}\includegraphics[width=1.3in, angle=-90]{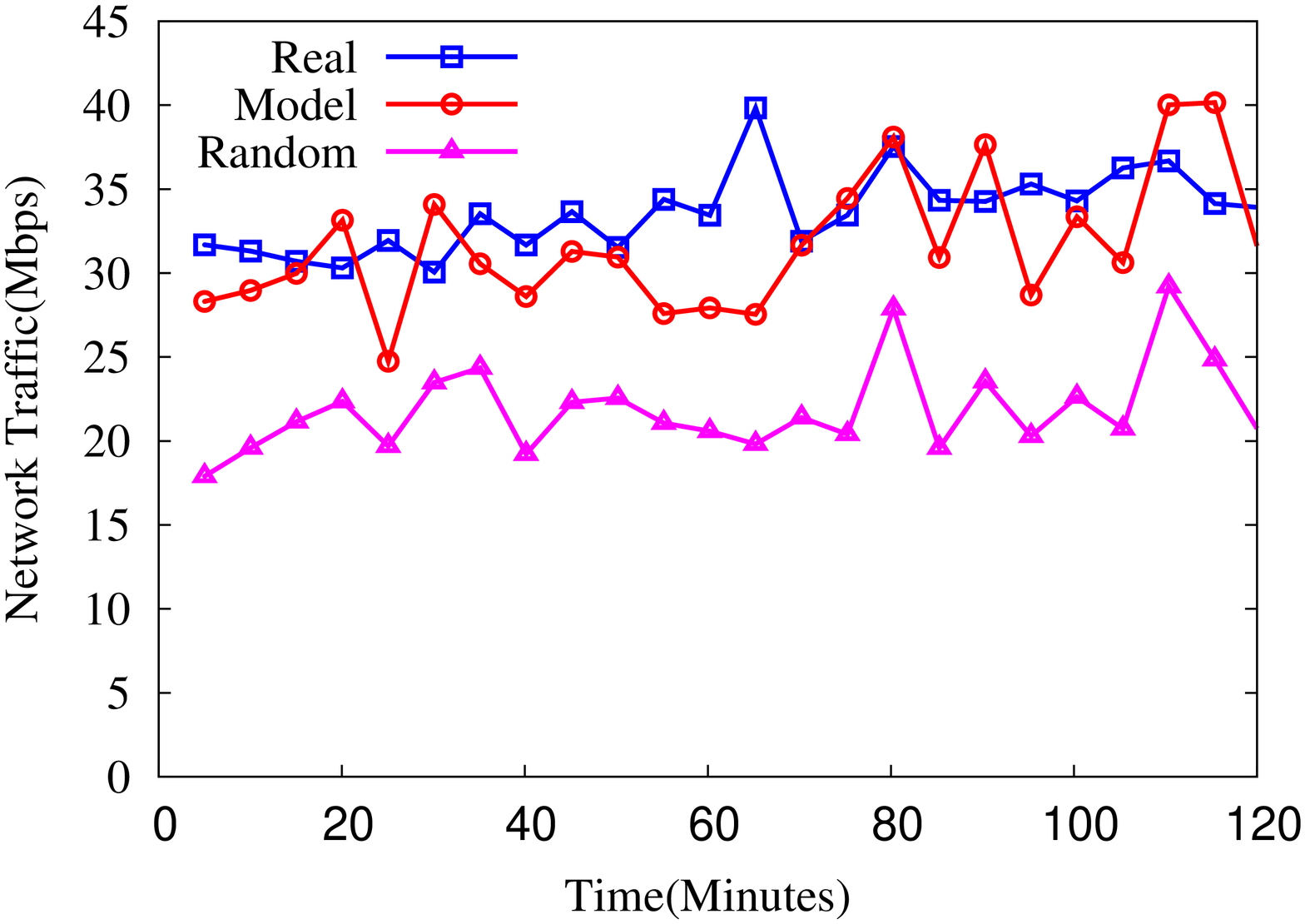}
&
\hspace{-0.7cm}\includegraphics[width=1.3in, angle=-90]{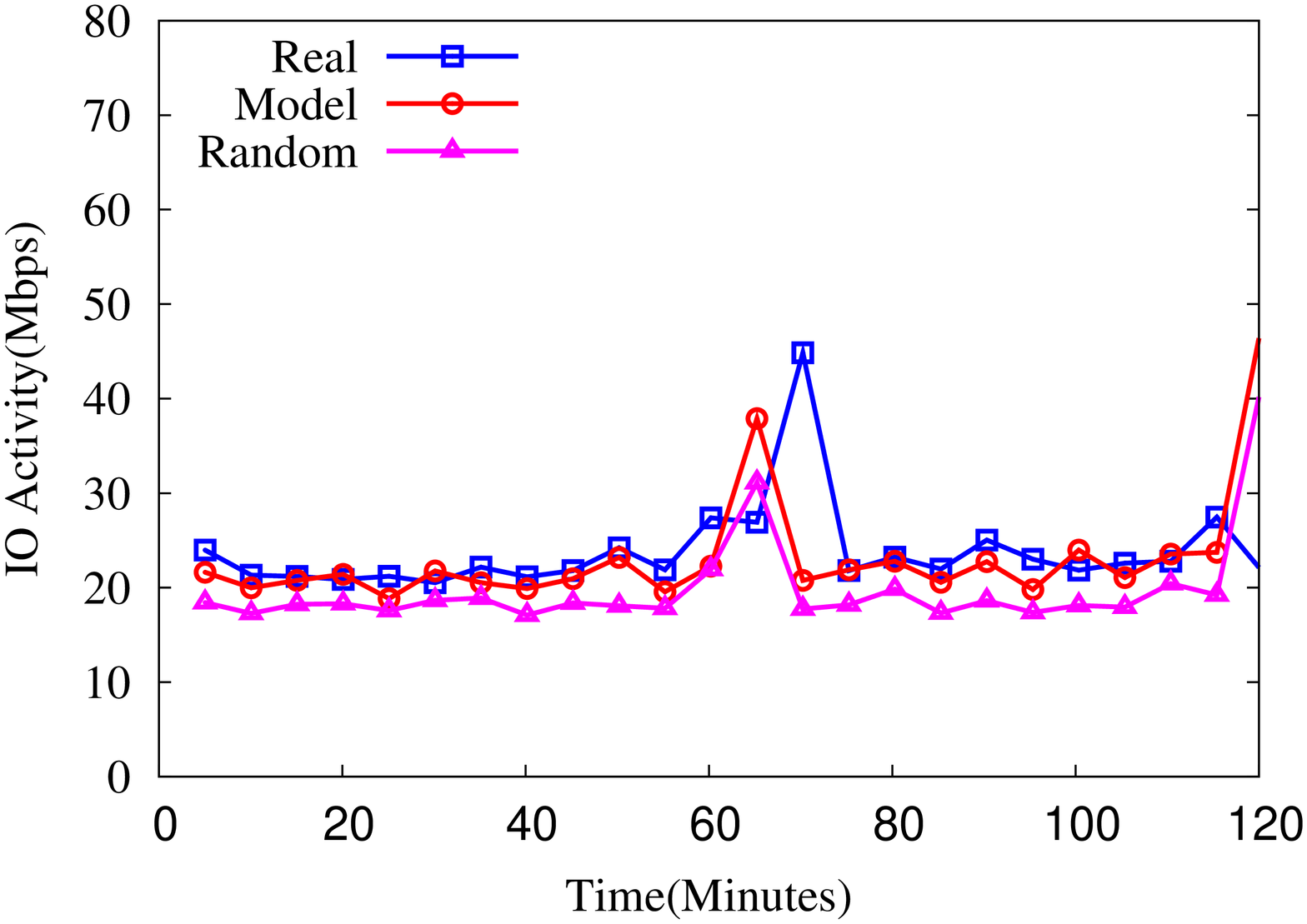}
&
\hspace{-0.7cm}\includegraphics[width=1.3in, angle=-90]{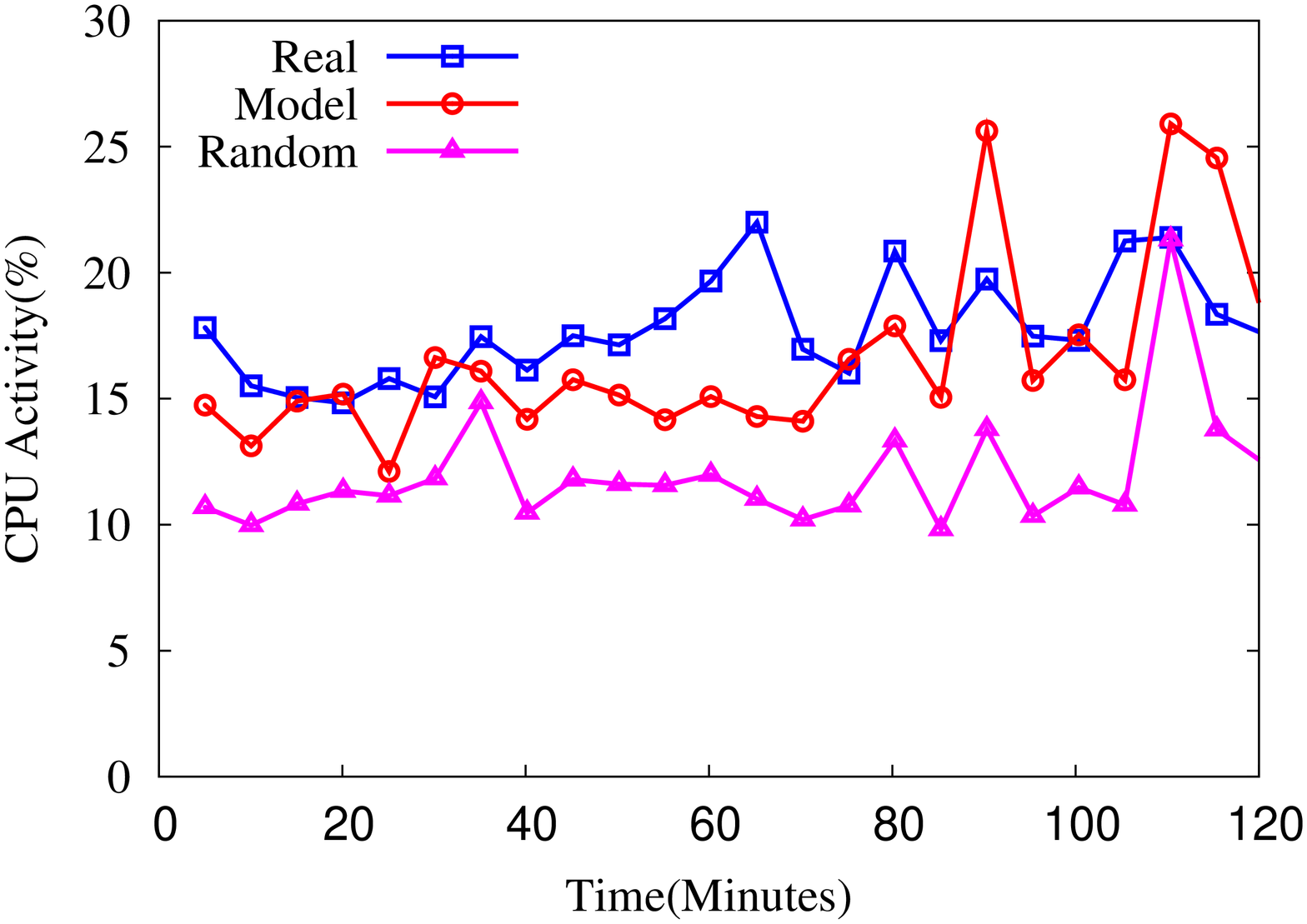}
\\
(a) & (b) & (c)
\end{tabular}
\vspace{-3mm}
\caption{ (a) Network traffic, (b) I/O activity, (c) CPU activity}
\label{fig:validate_model}
\end{figure*}

\subsection{Our OSN Implementation}

We implement a toy OSN that emulates most of the functionality of Twitter. 
In our system, each user can \textit{follow} other users by creating explicit links. Each user 
can update \textit{status} and can read the last 20 \textit{status updates} of all 
the \textit{one-hop} neighbors of the user. 
The back-end of our toy OSN system is implemented using Cassandra (ver 0.5.0) key-value datastore, with default settings 
and over 16 machines connected by a 1G Ethernet switch. Each machine is a Pentium Dual Core 2.33Ghz, 2GB of memory and 80GB of disk and run Ubuntu 8.04. The layout of how data is stored in our back-end is based on \cite{cassdata}. 

We load the social graph described in Sec.~\ref{sec:dat} into our system. 
For all the experiments, we use a dummy front-end python client that connects with Cassandra back-end 
through Thrift API. The front-end client parses the traces and directs requests to back-end by performing
Cassandra API calls. We use \textit{iostat} tool 
to monitor CPU and IO activity. For network traffic, we read \textit{/proc/net/dev/eth1} 
to access different counters of the network interface. 

\subsection{Validating the Model}

We first validate the model by generating realistic trace using SONG (using estimated parameters) of the busy-hour and comparing system parameters - network traffic in the back-end
, I/O activity and CPU activity against that of real data. In addition, we 
also compare against a trace that resembles the data in a temporal sense,
but the writes (tweets) are assigned uniformly at random to users
-- which is common practice. 

Fig.~\ref{fig:validate_model} shows the system activity with the 3 traces (each datapoint represents mean across 5 min bins). 
The random trace under-estimates the load 
across all the system parameters that we measured. In case of network traffic, the average traffic 
load is 36\% lower than the real trace. This result underscores the usefulness of using SONG to generate realistic traces.  
In contrast to the random model, the match between our trace and the real trace is much higher. For instance, 
the average network traffic only differ by 6\% among 2 traces. 
We like to note here that quantitative results are insignificant here, as one can tune Cassandra to optimize performance.
We are more interested in \emph{qualitative} results. 


\subsection{Benchmarking - Stress Test}

The next scenario we consider is a common task carried out by sysadmins  - to benchmark the system
and uncover the bottlenecks. One can do this by using benchmarking tools that are oblivious of the 
characteristics of application-specific workloads. However
as we showed in the previous section, using a trace with some grounding to reality can give more 
accurate results and can thus identify bottlenecks with higher confidence.

The typical steps followed for benchmarking include - generating a set of traces with 
increasing resource requirements, running the system with these different traces and monitor the QoS.  
Given a minimal target QoS, we can then determine the maximum workload that can be supported by the system infrastructure.

\begin{figure}[t]
\centering
\includegraphics[width=1.5in, angle=-90]{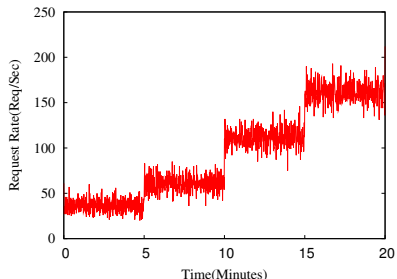}
\vspace{-3mm}
\caption{Tweets per seconds}
\label{fig:tweets_per_second}
\end{figure}

\begin{figure}[t]
\centering
\includegraphics[width=1.5in, angle=-90]{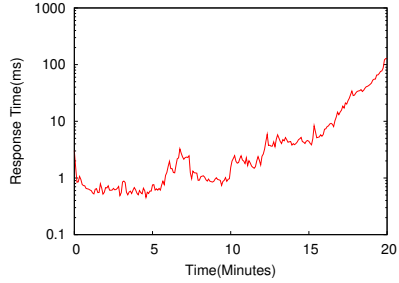}
\vspace{-3mm}
\caption{Response time across time}
\label{fig:response_time_across_time}
\end{figure}

We use SONG to generate traces by increasing the number of writes every 5 minutes 
from 25 writes/sec to 150 writes/sec (adding to existing writes, while keeping the variance and the overall write distribution across users 
the same as real data), using the extension discussed in Sec.~\ref{subsec:whatif}.
Fig.~\ref{fig:tweets_per_second} shows the write rate for 20 minutes in our test. We monitor 
the response time of every operation (get/put) of Cassandra from the front-end\footnote{Notice that 
the number of required Cassandra operations per new write depends to the number of followers of the user. }. 
Fig.~\ref{fig:response_time_across_time} show the average response time, in bins of 5 seconds. Notice that y-axis 
is in log scale. Despite variance in the response time, we observe that our Cassandra cluster is able to 
handle all the requests in less than 10 seconds, up to 100 writes/s.  When the request rate is about 150, our %
cluster collapses and the response time raise sharply up to more than 100 ms. 

\begin{figure*}[tbp]
\centering
\begin{tabular}[2]{ccc}
\hspace{-0.7cm}\includegraphics[width=1.3in, angle=-90]{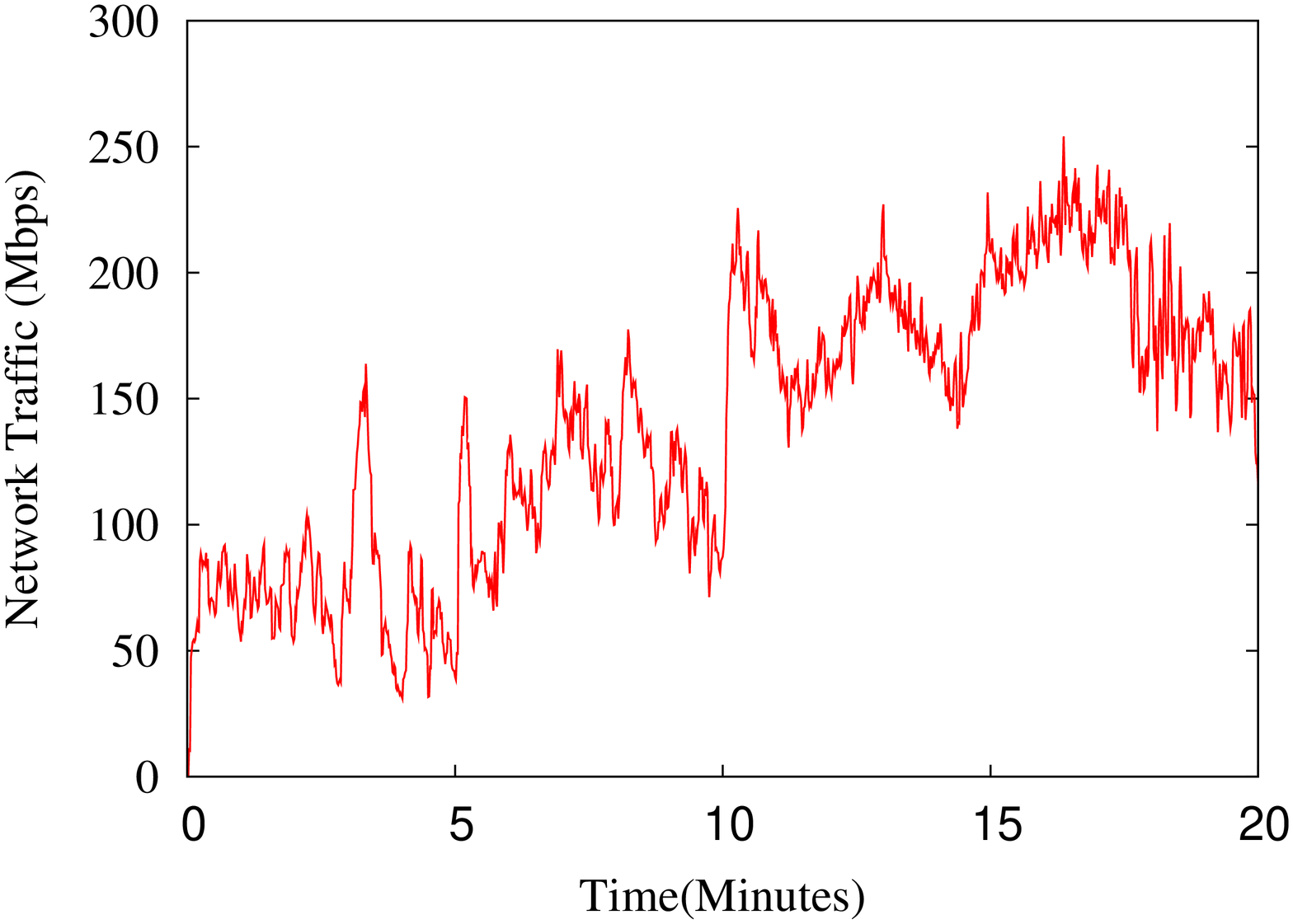}
&
\hspace{-0.7cm}\includegraphics[width=1.3in, angle=-90]{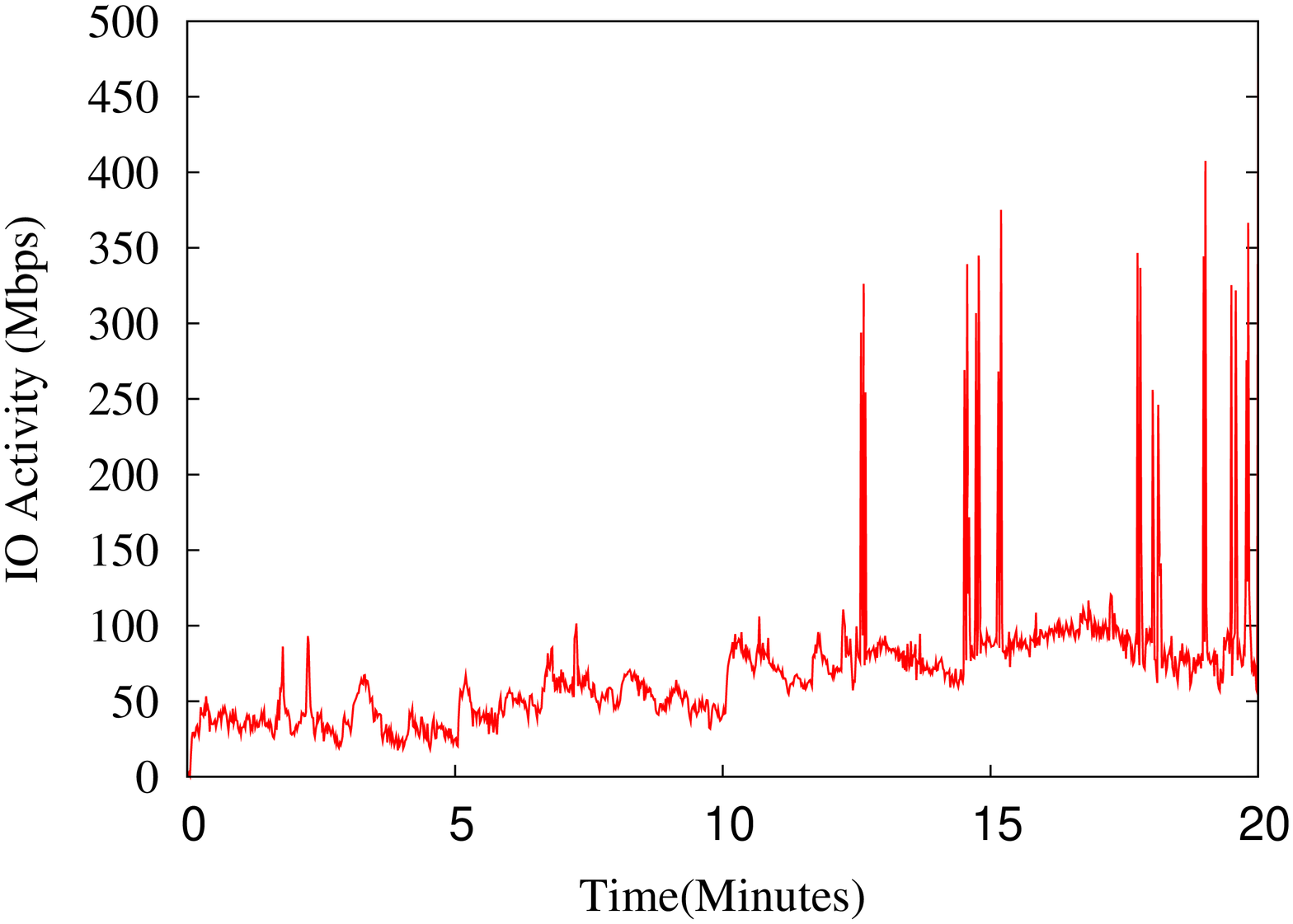}
&
\hspace{-0.7cm}\includegraphics[width=1.3in, angle=-90]{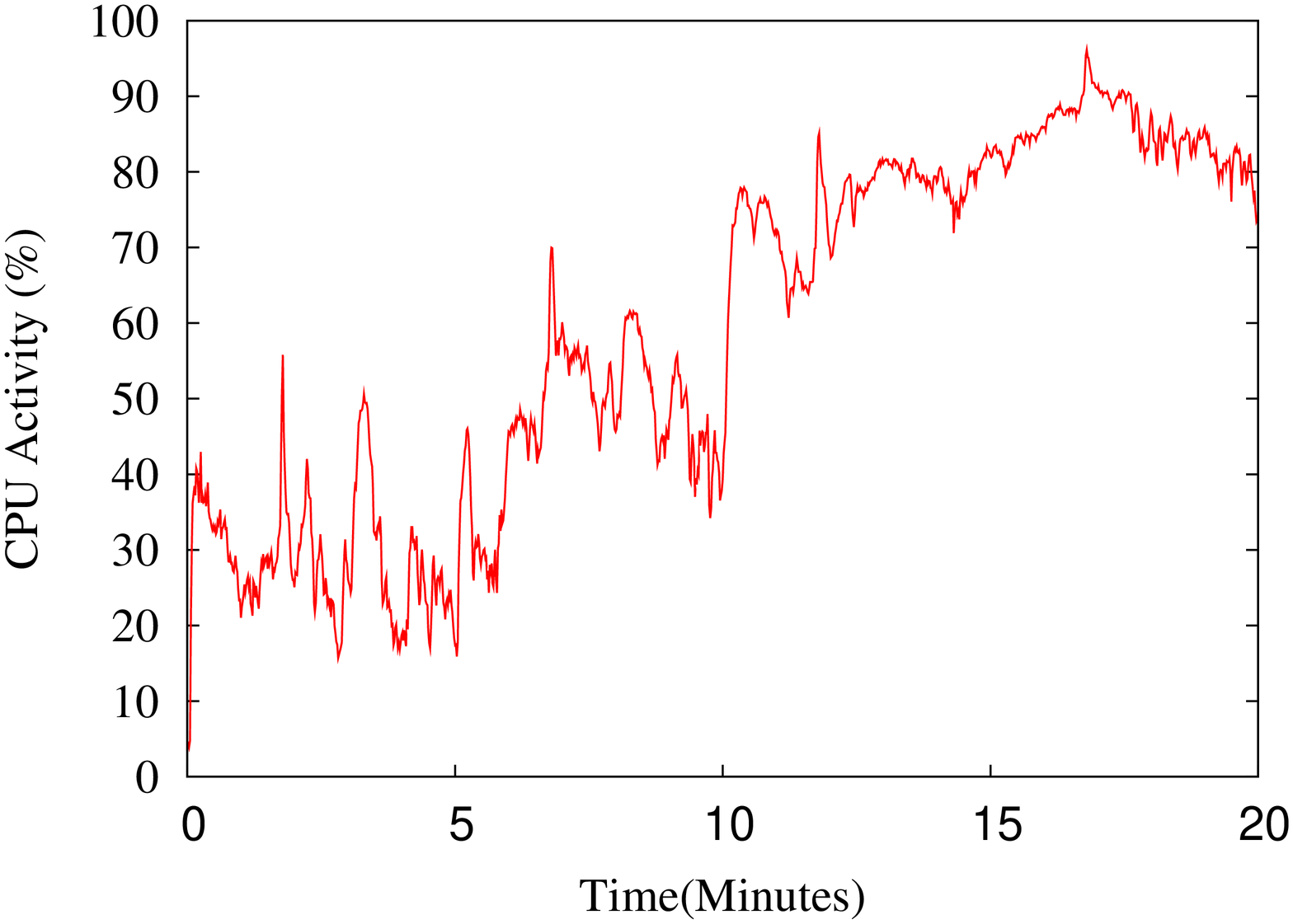}
\\
(a) & (b) & (c)
\end{tabular}
\caption{ (a) Network traffic, (b) I/O activity and (c) CPU activity}
\label{fig:stress-test}
\end{figure*}

Fig.~\ref{fig:stress-test} shows the activity in back-end network, I/O and the CPU utilization.  
We observe that although the total network traffic goes up to 250Mbps, 
we have a 1G switch, so network is clearly not the bottleneck.
We then checked the disk activity and it is only 40\% of the disk capacity.  
So we can conclude that CPU is our bottleneck that 
saturates the entire system when the load increases more than 100 tweets per second - as can be seen in Fig.~\ref{fig:stress-test} (c).
We'd like to reiterate again that this exercise is to show the usefulness and the versatility of SONG, as opposed
to uncover particular system issues with Cassandra. Hence the results should be interpreted qualitatively.


\section{Conclusions}
In order to conduct research in the increasingly popular area of OSNs, we need data-sets.
However, data-sets may be hard to get for a number of reasons. The main contribution of this paper
is SONG - a framework for generating synthetic and realistic traces of writes of users in OSNs. 
In order to develop the SONG framework, we characterized a large trace of write activity from Twitter.
We then developed the framework and showed how it can be used when prior traces are available or when
no traces are available. In order to show the effectiveness of SONG, we evaluated traces generated by SONG
using a real system implementation of a Twitter clone and showed by example the utility and versatility of SONG.
We intend to release code for SONG in the near future and hope it can be used by researchers and 
analysts
to generate traces to study different `what-if' scenarios.

\section{Acnowledgments}
Telefónica I+D participates in Torres Quevedo subprogram (MICINN),
cofinanced by the European Social Fund, for researcher's recruitment.

{\footnotesize \bibliographystyle{acm}
\bibliography{graph.bib}}


\end{document}